# Multiferroic Microstructure Created from Invariant Line Constraint

*Satyakam Kar*[a,b,c], *Yuki Ikeda*[d], *Kornelius Nielsch*[a,b], *Heiko Reith*[a], *Robert Maaß*[d,e,f], and *Sebastian Fähler*[c*]

[a]Leibniz IFW Dresden, Institute for Metallic Materials, 01069 Dresden, Germany

[b]TU Dresden, Institute of Materials Science, 01062 Dresden, Germany

[c]Helmholtz-Zentrum Dresden-Rossendorf, 01328 Dresden, Germany

[d]Federal Institute of Materials Research and Testing (BAM), 12205 Berlin, Germany

[e]Department of Materials Science and Engineering, University of Illinois Urbana-Champaign, 61801 Illinois, USA

[f]Department of Materials Engineering, Technical University of Munich, 85748 Garching, Germany

*Corresponding author, Tel.: +493512602775, e-mail: s.faehler@hzdr.de , Postal address: Helmholtz-Zentrum Dresden-Rossendorf, Bautzner Landstraße 400, 01328 Dresden, Germany

**Abstract**

Ferroic materials enable a multitude of emerging applications, and optimum functional properties are achieved when ferromagnetic and ferroelectric properties are coupled to a first-order ferroelastic transition. In bulk materials, this first-order transition involves an invariant habit plane, connecting coexisting phases: austenite and martensite. Theory predicts that this plane should converge to a line in thin films, but experimental evidence is missing. Here, we analyze the martensitic and magnetic microstructure of a freestanding epitaxial magnetic shape memory film. We show that the martensite microstructure is determined by an invariant line constraint using lattice parameters of both phases as the only input. This line constraint explains most of the observable features, which differ fundamentally from bulk and constrained films. Furthermore, this finite-size effect creates a remarkable checkerboard magnetic domain pattern through multiferroic coupling. Our findings highlight the decisive role of finite-size effects in multiferroics.

# 1. Introduction

Ferroic materials comprise ferromagnetic, ferroelectric, and ferroelastic materials, as well as combinations thereof, called multiferroics[1,2]. While the influence of finite sample size on ferromagnetic properties already fills several textbooks, finite-size effects in ferroelectrics[3,4] and multiferroics[5–7] are of current interest due to their high potential for miniaturized electronic devices. Of emerging interest are ferroelastic materials, which enable high-stroke high-force actuation[8–10]. They follow "the material is the machine" principle[11], which permits miniaturization due to a simple device design. However, further downscaling requires understanding finite-size effects during ferroelastic transitions.

Ni-Mn-based magnetic shape memory alloys are a promising multiferroic system exhibiting ferroelastic and ferromagnetic properties[12]. Coupling these two ferroic properties enables multifunctional applications suitable for microsystems like magnetic field-induced actuation[9,13,14], thermomagnetic waste heat harvesting[15,16], and multicaloric cooling[17]. Outstanding properties are obtained, when the first-order ferroelastic transition results in a steep, discontinuous change of magnetic order[12]. This ferroelastic transition is governed by a reversible martensitic transformation from a high-temperature austenite phase to a martensite phase at lower temperature. As this structural transformation is of first-order, lattice symmetry and parameters change discontinuously. At the transformation temperature, both phases coexist, and accordingly a compatible interface must form. This so-called 'habit plane' is nearly strain-free, which in most cases requires twinning of martensite[18]. In bulk, the crystallographic orientations of possible habit planes can be predicted just from lattice symmetries and parameters, as formulated first in 1950s[19,20] and further refined by Ball and James[21]. However, for freestanding thin films, Bhattacharya and James[22] proposed a theory different from bulk. As film thickness is much smaller than the film's lateral dimensions, the crystallographic compatibility conditions become more relaxed. Consequently, an 'invariant line' condition arises in thin films, which can allow the formation of an exact interface between (a) austenite and a single variant of martensite, and (b) two martensitic variants incompatible in bulk[23]. Thus, martensite microstructure in thin films should be fundamentally different from bulk, and deformation of films to tent and tunnel-like shapes were predicted[24,25]. Indeed, such a deformation has been shown experimentally in a single crystalline Cu-Al-Ni shape memory foil[11]. However, in all experiments on much thinner (500 – 2000 nm) constrained epitaxial Ni-Mn-Ga films[26–28], the habit plane concept known from bulk could explain the observed microstructure. To understand this discrepancy, we take a brief look at the few finite-size effects reported for epitaxial films. For films on substrates, the substrate is a rigid constraint for the martensite microstructure. To minimize the sum of twin boundary and elastic energies at this constraint, the twin boundary period scales as the square root of the film thickness[29]. For ultrathin films below 40 nm, this energy sum can even suppress the martensitic transformation completely[30]. Both examples illustrate that films on substrates are a particular case and significant differences are expected in freestanding films. Microstructure investigations of freestanding epitaxial films are sparse and do not provide a clear picture. They report slight[31] or no[32] changes after releasing the films. We attribute these ambiguous reports to the challenge of preparing and manually handling fragile freestanding films. This challenge has been solved recently by monolithic processing of epitaxially grown shape memory films on silicon, which utilizes standard silicon microtechnology to prepare partly freestanding films and patterns[33,34].

Using this approach here, we examine the martensite microstructure of a freestanding 500 nm thick epitaxial $Ni_{52}Mn_{18}Ga_{25}Cu_5$ film. Our analyses of complementary cross-sections by transmission electron microscopy (TEM) give a comprehensive 3D picture of the nested arrangement of different types of twin boundaries. Though these twin boundaries are *planes*, an invariant *line* constraint equation governs the key features of this complex microstructure.

Furthermore, this martensitic microstructure creates a unique checkerboard magnetic microstructure through multiferroic coupling. Our demonstration and analysis of the invariant line constraint enable understanding the finite-size effects in ferroelastic materials, which is critical for their applications in microsystems.

## 2. Results and discussion
### 2.1. From cross sections to the martensite microstructure in 3D

A martensitic microstructure is characterized by twin boundaries, which connect different orientations of martensite variants. Indeed, various types of twin boundaries are possible[35,36], and the resulting twin-within-twins hierarchy in constrained epitaxial Ni-Mn-Ga-based films has been explained recently[28]. In the present work, the as-deposited film exhibits this microstructure with an average mesoscopic twin boundary period of 117 ± 9 nm, as depicted in Supplementary Figure S1a. This microstructure changes fundamentally when the martensitic film is partially released from the substrate. As observed in Figure 1a, the surface of freestanding film exhibits two sets of orthogonal, parallel stripe-like features. A striking change in the scale of features is apparent in Figure 1b at the interface of freestanding and constrained portions of the film. The average period of features ($\Lambda$) increases about eight-fold to 994 ± 12 nm in the freestanding film.

To investigate this unique microstructure further, we prepared a TEM lamella across the interface, indicated by the white dashed line (CS1) in Figure 1b. We also introduce a (x, y, z) coordinate system aligned with $[100]_A$, $[010]_A$, and $[001]_A$ directions (A- austenite) to assign the features in 3D. An overview of the prepared cross-section TEM lamella is shown in Figure 1c, which highlights the constrained and freestanding regions, the silicon-on-insulator (SOI) substrate architecture, and three particular regions of interest in the film: freestanding (blue square), interface (red square), and constrained (green square). The freestanding region (Figure 1d) exhibits an unusual periodic array of triangle-shaped features arranged in an hourglass pattern, separated by rhombus-shaped features. These features contain a finer internal microstructure, analyzed in detail later. The rhombus features occur in two different thicknesses, alternating along the x-axis. These thickness variations are fundamentally different compared to the sawtooth surface topography of constrained films[31]. The cross-section of the constrained-freestanding interface in Figure 1e displays a sharp transition in microstructure, followed by the characteristic Type X martensite microstructure[26–28] in the constrained region (Figure 1f). Already at this rough scale, the cross-section images confirm the fundamentally different microstructure arrangement in the freestanding film, urging for a more in-depth investigation.

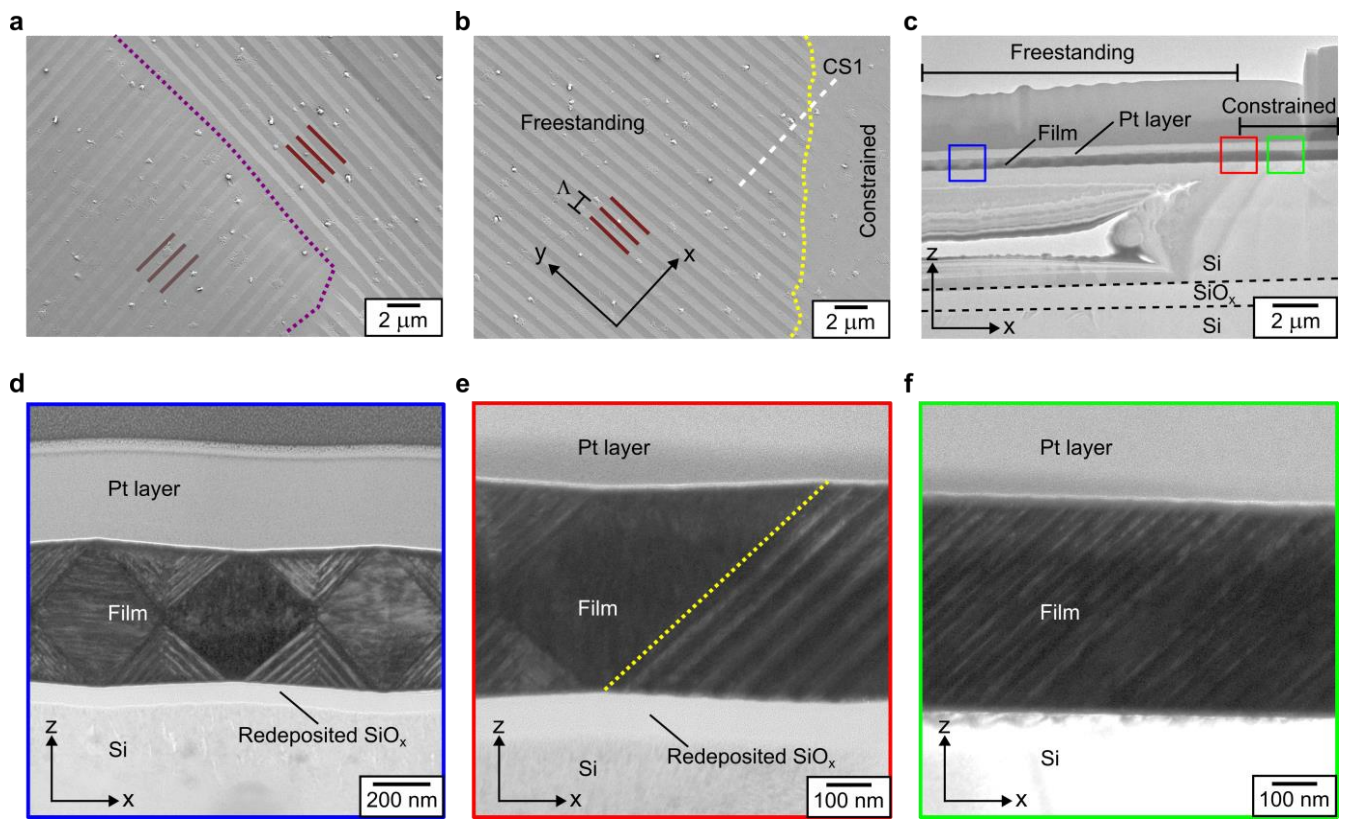

**Figure 1. Comparison between the freestanding and constrained martensite microstructure.** (a) Secondary electron image of the freestanding film surface shows parallel stripes aligned at 45° to substrate edges. Two equivalent stripe orientation (dark and light brown

lines) regions are observed, separated by a boundary (purple dotted line). (b) Secondary electron image of the freestanding-constrained interface reveals a sharp feature change at the interface (yellow dotted line). The stripes with a period Λ of about 1000 nm transition to features much smaller on the constrained side (see Supplementary Figure S1a). To examine this transition, FIB was used to prepare a film cross-section along the white dashed line (CS1). Its overview bright-field TEM image is shown in (c). Three key regions, marked by blue, green, and red squares, were investigated by bright-field TEM. (d) In the freestanding part, a periodic array of hourglass and rhombus features is observed – which is unique and further investigated in this paper. At the interface region in (e), this pattern transitions sharply to the well-known hierarchical martensite microstructure, which continues in the constrained region (f). The (x, y, z) coordinate system introduced here is used throughout the paper to assign the features within freestanding film. Note that etching by FIB redeposits $SiO_x$ below the freestanding part, an artifact analyzed in Supplementary Figure S2.

Therefore, we pursue a detailed TEM analysis of the unique hourglass and rhombus features in the x – z plane (Figure 2a). Locations A and B are marked inside the triangular features and differ in the orientation of boundaries visible inside. Nanobeam diffraction patterns of regions A and B reveal orthorhombic 14M martensite with $[010]_{14M}$ along the y-axis (see Supplementary Figure S3a for the exact location of diffraction). Two crystal orientations are visible in each diffraction pattern: one with $[100]_{14M}$ and the other with $[001]_{14M}$ aligned along the z-axis. This is the characteristic $a - c$ twinning of 14M martensite[37]. Hence, the parallel boundaries inside the triangular features are mesoscopic twin boundaries[28]. Locations marked C and D denote the thick and thin rhombuses in the film, with an average thickness of 523 nm and 461 nm, respectively. Additionally, a distinctly larger width of rhombus D than rhombus C is noticed along the x-axis. Each rhombus contains faint finely-spaced horizontal (C) and vertical (D) lines, indicating nano-twin boundaries. The selected area diffraction pattern of rhombus C reveals 14M martensite with $[010]_{14M}$ along the y-axis, while that of rhombus D reveals tetragonal NM martensite with $[010]_{NM}$ along the y-axis (see Supplementary Figure S3b for the exact location of diffraction). As per the adaptive concept[38], the 14M martensite can be described by a tetragonal NM martensite building block and periodic twinning at the nanoscale. Therefore, we stick to NM martensite to describe the rhombuses.

Using the obtained crystallographic information, we draw a simplified sketch in Figure 2b to mark the crystal orientations and label the identified boundaries in the x – z plane using the nomenclature from existing studies on bulk and constrained epitaxial films. The significant boundaries in this sketch include macroscopic boundary (separating triangular and rhombus regions), mesoscopic twin boundaries (inside the triangles), and nano-twin boundaries (inside the rhombuses). The mesoscopic twin orientations are close to $(101)_A$ and $(\bar{1}01)_A$ planes and conjugately related to each other[27,39]. Hence, the boundary separating the two conjugate twin orientations inside the triangles is called a conjugation boundary. The nano-twin boundaries observed in the rhombuses are close to $\{110\}_A$ orientation[39,40], and the four possible $\{110\}_A$ orientations (based on the observed traces in the x – z plane) are indicated by light and dark shades of green and blue dotted lines. This coloring scheme becomes more apparent in later investigations. The crystal orientation in the triangles alternates between $c_{14M}$ and $a_{14M}$ across the mesoscopic twin boundary and in case of the rhombuses, it switches between $a_{NM}$ and $c_{NM}$ across the nano-twin boundary. Although this schematic representation provides a good overview of the complex arrangement of boundaries in the martensite microstructure of freestanding film, their relative arrangement along the y-axis still remains to be explained.

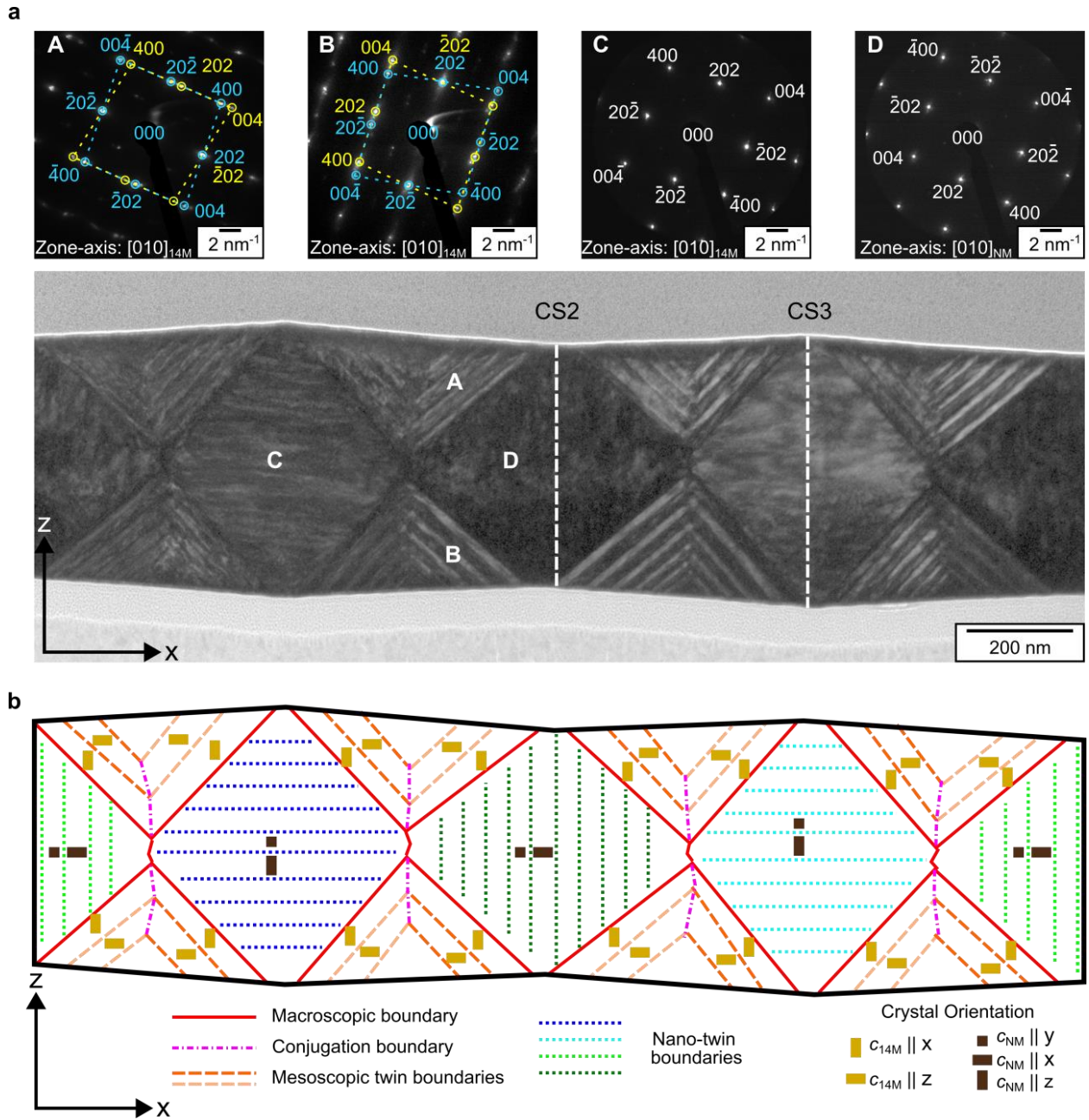

**Figure 2. Arrangement of martensitic variants and twin boundaries in cross-section CS1 across the stripe pattern.** (a) Bright-field TEM image reveals a periodic array of triangles (marked with A and B) arranged in hourglass feature. The hourglass features are separated by two types of rhombuses (marked C and D), which differ in thickness, width, and their internal finer features. Electron diffraction patterns, shown above the image, are used to identify crystal structure, their orientation (marked by their zone axis), and twin boundaries within these features. Triangles A and B consist of two 14M martensite variants (marked in blue and yellow),

which classifies the internal boundaries as $a - c$ mesoscopic twin boundaries. Rhombus C is indexed 14M martensite, whereas rhombus D is indexed NM martensite. The original lamella orientation used for these diffraction patterns is given in Supplementary Figure S3a,b. To obtain a 3D understanding of the microstructure in rhombuses, two additional cross-sections on positions equivalent to CS2 and CS3 are analyzed in Figure 3. (b) Sketch illustrating the boundaries and the crystal orientations observed in the microstructure. These intricate features are described in the main text.

To understand the microstructure within the rhombuses along the y-axis, we prepared two additional cross-section lamellae, CS2 and CS3, as indicated by white dashed lines in Figure 2a. Figure 3a shows the TEM analysis of CS2 cross-section in the $y - z$ plane. In the bright-field STEM image, we observe dark (E) and bright (F) regions, separated by boundaries. Both regions exhibit finely-spaced vertical lines similar to those observed in rhombus D in CS1 cross-section. The selected area diffraction patterns of regions E and F reveal 14M martensite with $[100]_{14M}$ along the x-axis (see Supplementary Figure S3c for the exact location of diffraction). Interestingly, another diffraction pattern in this cross-section reveals NM martensite with $[001]_{NM}$ along the x-axis (Supplementary Figure S4). Hence, these diffraction patterns indicate that the nano-twin boundaries have coarsened in some parts of the rhombus, leading to the coexistence of 14M and NM martensite[37]. Based on these observations in CS2 cross-section, we sketched the microstructural features and crystal orientations in Figure 3b. The nano-twin boundaries fill most of the cross-section and are oriented vertically, both in CS1 (rhombus D) and CS2 cross-sections. From the two traces of each twin plane, we derive their orientations as $(110)_A$ and $(\bar{1}10)_A$. Although these twins are crystallographically equivalent, their orientation difference causes a slight contrast between regions E and F. The orientations are conjugately related, and hence, the two regions are separated by a conjugation boundary.

The crystal orientations in both regions alternate between short $a_{NM}$ and long $c_{NM}$ lattice parameters across the nano-twin boundary. Thus, rhombus D contains two different orientations of nano-twin boundaries along the y-axis, which make vertical traces in cross-sections CS2 and CS1.

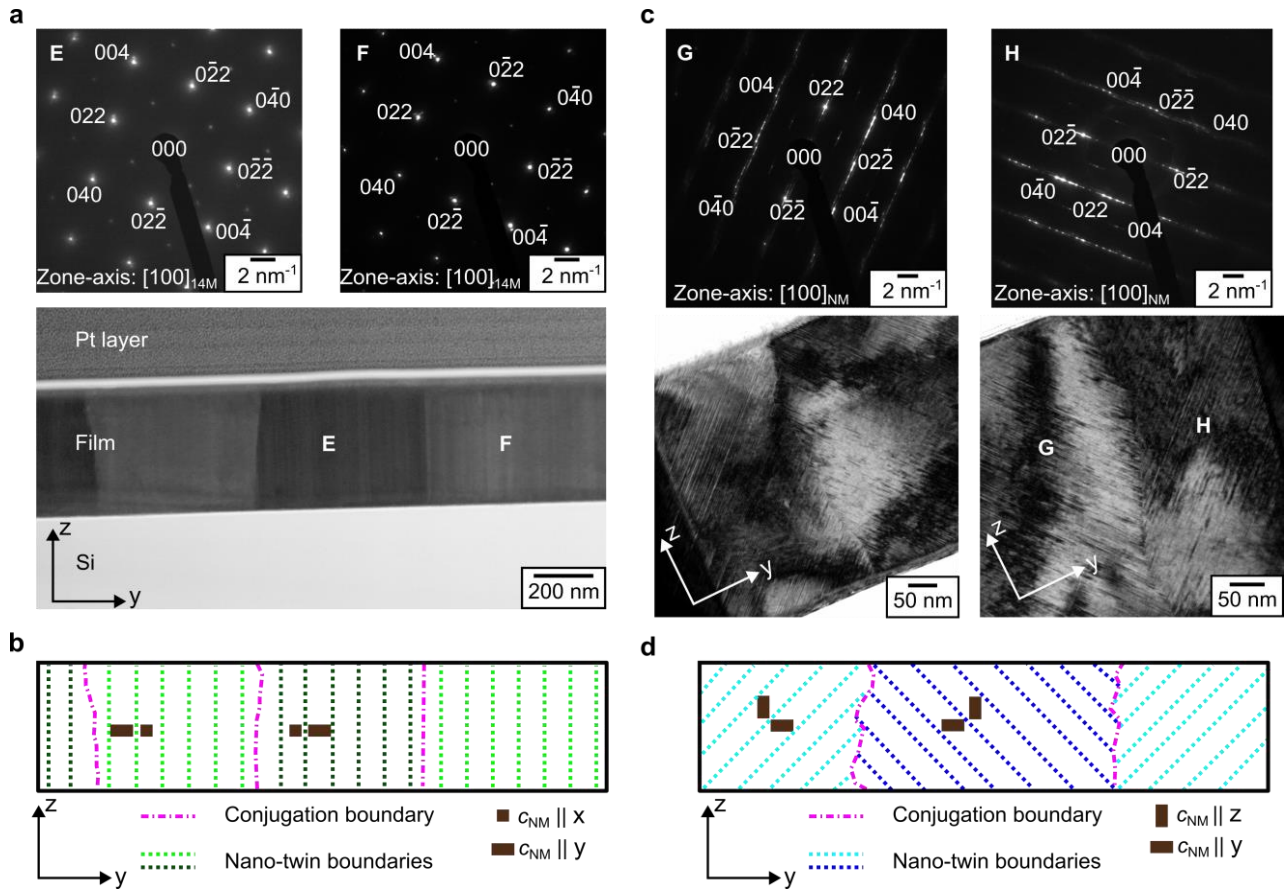

**Figure 3. Martensite microstructure within cross-sections CS2 and CS3 along the stripe pattern.** (a) Bright-field STEM image of a CS2 cross-section (rhombus D) exhibits finely-spaced vertical lines and two unique regions (marked E and F) with a contrast. Selected area diffraction patterns from E and F, shown above the image, agree with 14M martensite. The original lamella orientation used for these diffraction patterns is given in Supplementary Figure S3c. (b) Sketch illustrating the boundaries and crystal orientations observed in (a). (c) Bright-field TEM images of a CS3 cross-section (rhombus C) show two regions (marked G and H) containing finely-spaced diagonal lines oriented at about 45° to the z-axis. Selected area

diffraction patterns from the two regions, shown above the image, agree with NM martensite in $[100]_{NM}$ zone-axis. In addition, satellite spots along the $g_{022}$ vector in the diffraction pattern G and $g_{0\bar{2}2}$ vector in the diffraction pattern H are noticed, indicating the presence of 14M martensite as well. (d) Sketch illustrating the boundaries and crystal orientations observed in (c).

The same analysis is repeated for rhombus C by analyzing CS3 cross-section in the y – z plane (Figure 3c). The microstructure in the bright-field TEM images reveals finely-spaced lines aligned at 45°. Two different alignments with respect to the z-axis are possible, and accordingly, one can separate the two regions (marked G and H) on either side of a wiggly boundary. The selected area diffraction patterns of the regions G and H reveal NM martensite with $[100]_{NM}$ along the x-axis. Furthermore, these diffraction patterns contain satellite spots spaced with $(\frac{1}{7})g_{022}$ between the main spots[41]. This implies that in regions G and H, 14M martensite is also present, similar to the cross-section CS2. Based on the observed features, the CS3 cross-section is sketched in Figure 3d. The nano-twin boundaries fill regions G and H, and the trace analysis in CS1 (rhombus C) and CS3 cross-sections identify their orientations as $(011)_A$ and $(0\bar{1}1)_A$. As both twin orientations are also conjugately related, we separate the two regions by a conjugation boundary. The crystal orientation is similar to CS2 cross-section, with alternating short $a_{NM}$ and long $c_{NM}$ lattice parameters across the nano-twin boundary. Thus, rhombus C also contains two different orientations nano-twin boundaries along the y-axis separated by conjugation boundaries. Furthermore, 14M and NM martensite coexist in both rhombuses, which indicates a coarsening of the nano-twin boundaries. A coarsening of the constrained martensite microstructure is also noticed at the interface between constrained and freestanding portions in the y – z plane, analyzed in Supplementary Figure S5.

The complete martensitic microstructure is summarized as a 3D sketch in Figure 4. What appears as a simple stripe pattern at the surface (Figure 1a) is indeed much more complex below the surface. Although these boundaries are known from bulk, the origin for their nested arrangement remains to be explained.

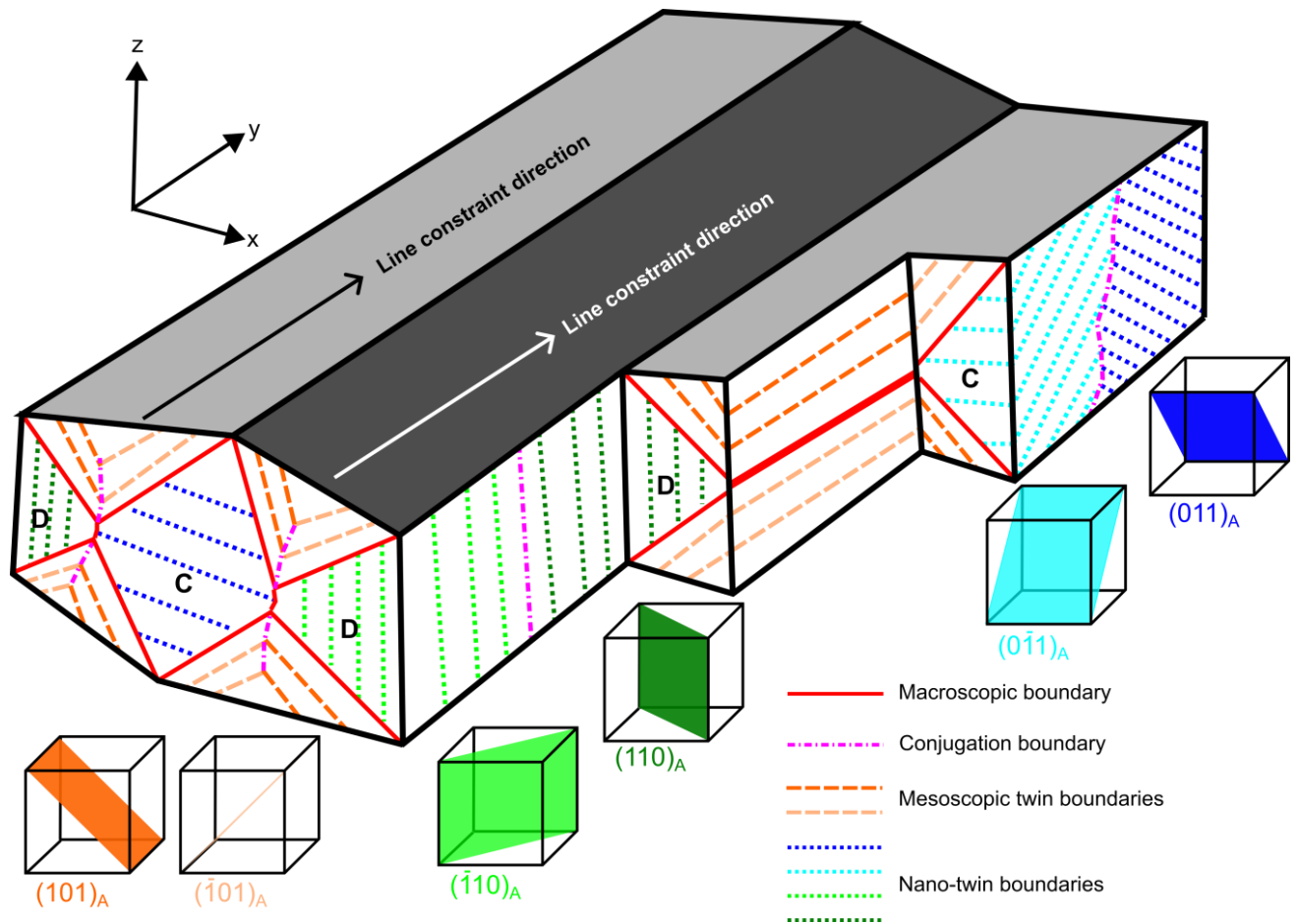

**Figure 4. 3D sketch of the martensite microstructure in freestanding film.** The combination of different cross-sections allows assembling the nested arrangement of different types of boundaries. We explain in Section 2.2 that the observed unique microstructure is a consequence of an invariant line constraint along the y-axis.

**2.2. Explaining the microstructure by a line constraint**

To understand the origin of this peculiar martensite microstructure, we consider the general concept that a martensite microstructure is guided by minimizing elastic energy through the introduction of compatible twin boundaries. In bulk continuum theory, this constraint results in two matrix equations, which use the lattice parameters of austenite and martensite as input and define the possible orientations of habit and twin planes[19,20]. For the present case of freestanding film with a finite thickness, we will show now that this *plane* constraint reduces to a *line* constraint along the stripe features, and accordingly we obtain a simple linear equation.

We start with the key experimental observation that no macroscopic strain occurs in the x – y plane of the freestanding film and at the connection to the constrained part of the film (Figure 1b). Consequently, the in-plane film dimensions, determined by the austenite state during film deposition, remain invariant during the martensitic transformation and subsequent film release process. To fulfill this invariant constraint along the y-axis in the rhombus features, an alternating arrangement of tetragonal NM martensite unit cells, with short ($a_{NM}$) and long ($c_{NM}$) lattice parameters, is needed to match the length of the same number of austenite unit cells with lattice parameter $a_A$. This alternating arrangement must occur in a *fine mixture* ($\lambda_Y$), such that:

$$a_A = \lambda_Y \times c_{NM} + (1 - \lambda_Y) \times a_{NM} \qquad (\text{eq. 1})$$

Using the particular lattice parameters for our film characterized by XRD (see supplementary Figure S1b, Table S1), we obtain $\lambda_Y = 0.313$. This alternating arrangement of unit cells is realized by nano-twin boundaries known from bulk. As these boundaries have a very low twin boundary energy[38], they can be introduced at spacings much shorter than film thickness. This microstructure with twin *planes* is observed within the rhombuses and highlighted in the sketch (Figure 4) by shades of green and blue dotted lines.

An invariant line constraint along the y-axis has consequences on the extension of the rhombuses along the z- and x- axes. For instance, all unit cells in rhombus D are aligned with $a_{NM}$ along the z-axis (see Figure 2). As $a_{NM}$ is shorter than $a_A$, rhombus D decreases in thickness, in agreement with experimental observation. To make up for volume conservation, the rhombus expands along the x-axis. This modified length ($l$) relative to $a_A$ along the x-axis is also determined by $\lambda_Y$, as per the equation given in Table 1 for $l_X(D)$.

As the overall film extension along the x-axis remains invariant, the neighboring rhombus (rhombus C) must have a shorter extension along x-axis and consequently elongate along the z-axis to conserve volume. This explains the experimentally observed alternating arrangement of thick rhombus C and thin rhombus D along the x-axis. Additionally, we performed a quantitative comparison of the calculated thickness and width ratios of the rhombuses with that observed in cross-section CS1. These values along with the used equations are summarized in Table 1. The good agreement between calculated and experimentally measured values strongly supports the invariant line constraint concept. It is worth mentioning that our equations neglect tilting of lattice, which additionally occurs at a twin boundary. As worked out in Supplementary Table S2, it has little effect on the calculated values. Tilting also causes a shear stress, which is not incorporated within our model. Following the work of Roytburd[36], this shear deformation is diminished at a larger length scale by conjugation boundaries, which connect two conjugate orientations of nano-twin boundary along the y-axis (Figure 4).

To understand the origin of triangles forming the hourglass feature, we assume in a thought experiment that they are absent. Thus, at the connection between thick rhombus C and thin rhombus D, a step would appear at the surface. To avoid this step, a huge elastic deformation within both rhombuses would be required. Instead of a step, it is energetically more favorable to introduce triangles as connectors, which consist of a martensitic microstructure well known from constrained films[26–28]. The triangles are slightly rotated by 3° with respect to the average film surface to compensate for the different thicknesses of rhombuses C and D. Triangles also fulfill the invariant line constraint by aligning $b_{14M}$ along the y-axis as per the adaptive concept[38].

Finally, we consider the consequences of the invariant line constraint at much larger length scales. A line constraint acts ad-infinitum, which explains the long parallel stripes on the film surface, reaching several tens of micrometers. The parallel stripes end when they meet the other orthogonal stripe orientation (Figure 1a), which is crystallographically equivalent due to the four-fold in-plane symmetry of the epitaxial film. In summary, the invariant line constraint uses twin planes known from bulk to successfully explain most features observed in the martensite microstructure of freestanding films.

**Table 1**. Comparison between the experimentally measured and calculated values of the rhombus features in the freestanding film microstructure. The equations used for the calculations are included here and described in the main text.

| Rhombus feature | Measured value | Calculated value | Equations used |
|---|---|---|---|
| Ratio of thickness (C : D) | 1.133 ± 0.0063 | 1.140 | $l_Z(C) = \lambda_Y \times a_{NM} + (1 - \lambda_Y) \times c_{NM}$<br>$l_Z(D) = a_{NM}$<br>Ratio $= l_Z(C)/l_Z(D)$ |
| Ratio of width (D : C) | 1.136 ± 0.0071 | 1.140 | $l_X(D) = \lambda_Y \times a_{NM} + (1 - \lambda_Y) \times c_{NM}$<br>$l_X(C) = a_{NM}$<br>Ratio $= l_X(D)/l_X(C)$ |

**2.3. Anisotropic checkerboard magnetic microstructure**

Magnetic shape memory alloys are multiferroic in nature, where the magnetic and martensitic microstructures are coupled[12]. Hence, we use magnetic force microscopy (MFM) to analyze the domain pattern of the freestanding film (Figure 5). The simultaneously probed surface topography allows for a correlation with the features of martensite microstructure, consisting of alternating thick and thin rhombuses (Figure 5a). Its corresponding magnetic microstructure reveals a peculiar checkerboard-like pattern (Figure 5b). This pattern is fundamentally different from the characteristic band domain pattern of constrained films[42], seen in Supplementary Figure S6. The observed pattern is also anisotropic as it differs along the x- and y-axes. The average domain width is 357 ± 39 nm along the y-axis and 275 ± 17 nm along the x-axis. Further differences are best visible from two perpendicular line profiles, $L_X$ and $L_Y$, marked in Figure 5a, b. The extracted line profile $L_Y$ from the phase signal shows an irregular peak to peak spacing (Figure 5c), with no correlation to the surface topography. In contrast, perpendicular to the stripes along $L_X$ (Figure 5d), the phase signal appears strongly correlated with the periodic peak and valley topography. The phase contrast inverts from bright to dark (or vice-versa) across a peak and about midway between each peak and valley (where the triangles are located) but remains unaffected across a valley. This is also evident from the overlay of magnetic microstructure on topography of the scanned region (Supplementary Figure S7). The coupling of magnetic and martensite microstructure along the x-axis is sketched in Figure 5e, which highlights the locations of contrast inversion and mirroring with respect to the surface topography.

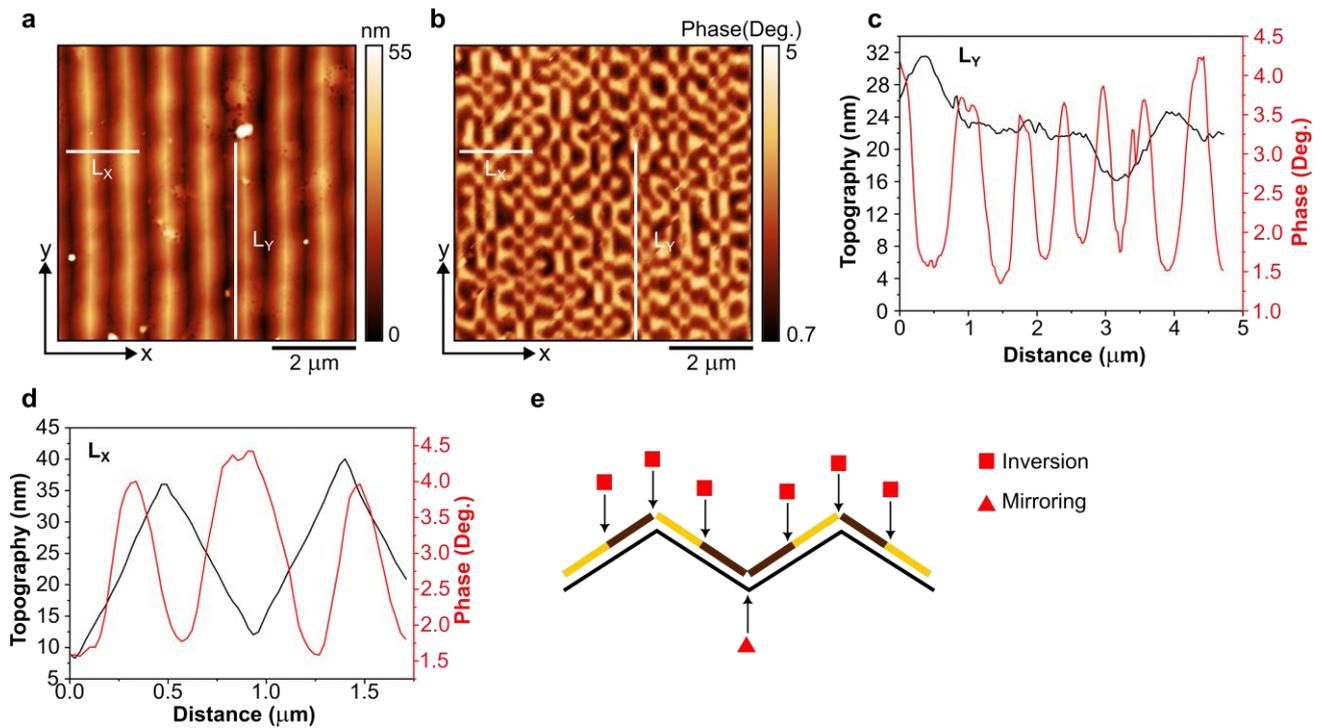

**Figure 5. Anisotropic checkerboard magnetic microstructure on the surface of freestanding film.** (a) Surface topography of the freestanding film and its (b) corresponding magnetic microstructure obtained by magnetic force microscopy. Line profiles $L_X$ and $L_Y$ (marked in white) are extracted perpendicular and parallel to the stripe pattern, respectively. (c) Parallel to the stripes ($L_Y$), changes of phase contrast are irregular (c) Perpendicular to the stripes ($L_X$), a complex coupling occurs between phase contrast and topography. The phase contrast changes from dark to bright (or vice-versa) at the peaks and mid-way between peak and valley but remains unchanged across a valley. (e) Sketch illustrating the observed coupling along the x-axis. The locations of contrast inversion are highlighted (peaks and midway between peaks and valleys). In addition, the sequence of contrast is mirrored at valleys.

We can derive all features of this unique magnetic microstructure by considering the multiferroic coupling, given by magnetocrystalline anisotropy and magnetostatic energy minimization, essential for films with finite thickness. Using the crystallographic information from Figure 2b, the magnetic easy-axis orientations are sketched for CS1 cross-section in Figure 6a. We use the easy-axes known from bulk, which differ for orthorhombic 14M martensite, having a magnetic easy-axis along $c_{14M}$[43], and tetragonal NM martensite, having an easy plane of magnetization with two easy-axes along $a_{NM}$[44]. The solution for the magnetization directions must consider that magnetization follows the easy-axis to minimize anisotropy energy, and charged domain boundaries are avoided to minimize magnetostatic energy. One solution of the magnetic domain configuration, which fulfills these conditions everywhere, is sketched in Figure 6b. As each easy-*axis* allows for two alignments of the magnetization *directions*, another equivalent solution is the domain arrangement, where all magnetization directions are reversed. We consider these two solutions optimum, as they completely avoid charged domain boundaries within the complex hourglass feature and require just one horizontal 180° domain wall in the center of the thick rhombus. Furthermore, these solutions explain the experimentally observed occurrences of contrast inversion and mirroring along the x-axis sketched in Figure 5e. This contrast inversion originates from the arrangement of mesoscopic twin boundaries, which act like mirrors for the magnetic domains. This effect is known from previous experiments in Ni-Mn-Ga bulk single crystal[45] and constrained epitaxial films[32,42].

In the proposed solutions, some regions do have magnetization components along the z-axis, which results in an unfavorable stray field coming out of the sample surface. In order to minimize this magnetostatic energy, an alternating arrangement of both solutions occurs along the y-axis, connected by additional 180° domain walls. The balance of magnetostatic and domain wall energy is a common mechanism for domain formation in films with out-of-plane easy-axis. The resulting magnetic microstructure in the x – y plane is sketched in Figure 6c, with yellow and brown shades for the out-of-plane component, which is probed by MFM. This coloring scheme assumes that due to the small twinning period in the film, magnetic exchange energy tends to align magnetization in parallel and thus short alternations of easy-axis are not resolvable by MFM. To sum up, our phenomenological model explains all features of the peculiar magnetic microstructure based on the underlying martensite microstructure. Moreover, this line of reasoning illustrates that at these length scales, the martensite microstructure governs the magnetic one, and not vice-versa. We attribute this to the relevant interfacial (twin boundaries / domain walls) and volume energies (elastic / magnetostatic), which are much higher for the ferroelastic than the ferromagnetic subsystem[29,42] and thus must be minimized first.

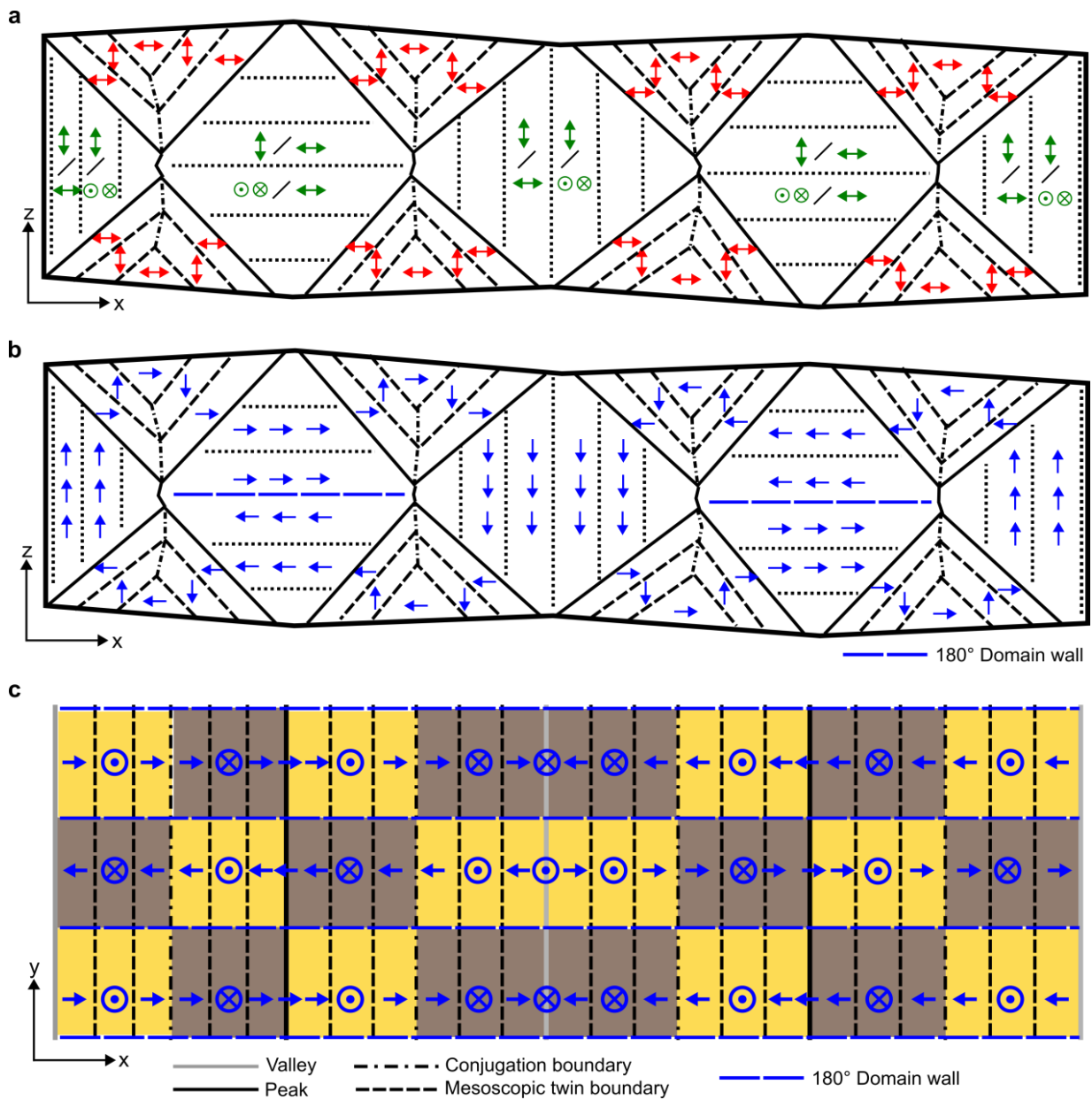

**Figure 6. Phenomenological explanation of the anisotropic checkerboard magnetic microstructure.** (a) Starting point is the crystal orientation shown in Figure 2b, which determines the magnetic easy-axis orientations, sketched here for CS1 cross-section. (b) Magnetization direction must follow these easy-axis orientations and magnetically charged interfaces must be avoided, which is achieved by the sketched magnetic domain configuration, requiring just one 180° domain wall in the center of each thick rhombus. Another equivalent solution is reversing magnetization directions of domains altogether. The domain configuration within the film determines the magnetic domain arrangement on the film surface, sketched in (c). Along the y-axis, the alternating arrangement of both equivalent solutions minimizes the magnetostatic energy. Yellow and brown colored regions mimic the magnetic microstructure. Key features of the martensite microstructure are overlaid to illustrate that the unique features of this domain pattern are determined by the line constraint.

In summary, a size reduction from bulk to films changes the constraint of a martensitic transformation from an invariant plane to an invariant line. The line constraint gives rise to simple linear equations that quantitatively explain most of the observed features of the twinned martensite microstructure. This microstructure is unique for freestanding films, as other geometries do not fulfill the required preconditions. For films on substrates, the rigid constraint inhibits the formation of the rhombus feature, and in bulk, no dimension is thin enough to allow for the thickness variation associated with the rhombus feature. In ultrathin films, twin boundaries are finally expected to converge to a twin line; thus, the twin boundaries required here cannot exist. Hence, the line constraint is decisive in freestanding films with finite thickness – the geometry favorable for most applications described in the introduction. This line constraint also determines the magnetic microstructure of our multiferroic films.

Therefore, we anticipate that the finite-size effect uncovered here will be of high relevance for a variety of emerging multiferroic membranes with reduced dimensionality[46,47].

## 3. Methods

3.1. Film deposition

A 500 nm thick epitaxial Ni-Mn-Ga-Cu film was grown on commercially available 4 nm $SrTiO_3$ (STO) buffered silicon-on-insulator (SOI) substrates by DC magnetron sputtering. This approach was demonstrated in our previous work[33]. The film was prepared by co-sputtering a $Ni_{48}Mn_{22}Ga_{30}$ alloy target and an elemental Cu target on substrates heated to 673 K.

3.2. Film release process

The deposited film was partly released from the substrate by selectively etching parts of the film, followed by under-etching the top Si layer of the SOI substrate. The film was patterned by laser photolithography and Ar ion beam etching. The top Si layer of the SOI substrate was selectively under-etched by $XeF_2$ gas etching. The process details of lithography, Ar ion beam etching, and $XeF_2$ etching are mentioned in our previous work[34].

3.3. Film Characterization

The crystal structure of the as-deposited film was analyzed using a Bruker AXS D8 Advance X-ray diffractometer with Co-radiation ($\lambda_{K\alpha}$ = 0.178897 nm). A *θ - 2θ* scan in Bragg-Brentano geometry with sample tilt offsets (Δω) from 0° to 10° (step size = 1°) was used to account for the slight tilt of crystal planes of martensite[48]. The film contains both 14M martensite and NM martensite at room temperature. (Supplementary Figure S1b). Iso-field magnetization measurement as a function of temperature was performed on the as-deposited film in a vibrating sample magnetometer (Quantum Design-Versalab) with an applied field of 0.1 T in-plane along the $[110]_A$ direction and in the temperature range of 150 – 400 K and scan rate of 0.033 K s$^{-1}$. The Curie transition temperature is measured to be 347 K (Supplementary Figure S1c). The film microstructure and composition were analyzed in a ZEISS Sigma 300 scanning electron microscope fitted with a SmartEDX detector. Using a bulk $Ni_2MnGa$ standard, the composition of the films was determined with 1 at. % accuracy to be $Ni_{52}Mn_{18}Ga_{25}Cu_5$. An FEI Quanta 3D FEG device and an FEI Helios Nanolab 600i were used to prepare freestanding film cross-section lamellae by focused ion beam (FIB) etching technique. The TEM investigations on the lamellae were carried out in an FEI Talos F200S and a JEM-2200FS transmission electron microscopes operated at 200 kV. The electron diffraction patterns were analyzed using the CrysTBox software[49]. All electron diffraction patterns from TEM analyses are indexed using the $L2_1$ convention[50]. Magnetic force microscopy of the films was performed on a Bruker Dimension Icon atomic force microscope with a two-pass LiftMode™ technique. An MESP-V2 tip magnetized along its axis was used with a lift height of 100 nm. The obtained height sensor data were used to image the surface topography and the phase data to image the magnetic stray field of the sample.

**Acknowledgements**

This research was performed under a BAM-IFW tandem project. The authors acknowledge Almut Pöhl (IFW) for preparing a TEM lamella and Carsten Prinz (BAM) for support with TEM investigation. The authors acknowledge Adnan Fareed (BAM), Klara Lünser (UB), and Volker Neu (IFW) for discussion. The TEM investigations were carried out at the electron microscopy center at BAM.

**Author contributions**

 S.F. and S.K. designed the study. H.R., K.N., and R.M. provided the resources. Y.I. performed the TEM observations. S.F. proposed the line constraint concept. S.K. performed the remaining experiments, analyses, and wrote the first version of the manuscript. All authors contributed to the final version.

**Competing interests.** The authors declare no competing interests.

**Additional information.** Supplementary information is linked to the online version of the paper.

**Data and materials availability.** Raw data of all experiments supporting this study are available at RODARE (DOI:10.14278/rodare.2889)